\documentstyle[preprint,aps]{revtex}
\tightenlines

\begin{document}
\newcommand{\beq}{\begin{equation}}
\newcommand{\eeq}{\end{equation}}
\newcommand{\beqa}{\begin{eqnarray}}
\newcommand{\eeqa}{\end{eqnarray}}
\newcommand{\sr}{\sqrt}
\newcommand{\fr}{\frac}
\newcommand{\mn}{\mu \nu}
\newcommand{\G}{\Gamma}

\draft
\preprint{ INJE-TP-00-07, hep-th/0010208 }
\title{No-go theorem for  gravivector  and graviscalar   \\
       on the brane}
\author{  Y.S. Myung\footnote{E-mail address:
ysmyung@physics.inje.ac.kr} }
\address{
Department of Physics, Graduate School, Inje University,
Kimhae 621-749, Korea}
\maketitle

\begin{abstract}
We prove the no-go theorem that the gravivector $(h_{5\mu})$ and graviscalar
$(h_{55})$
cannot have any on-shell propagation on the  Randall-Sundrum (RS) brane.
For this purpose, we analyze all of their linearized equations
 with the de Donder gauge (5D transverse-tracefree gauge).
 But we do not introduce any matter source.
We use  the Z$_2$-symmetry argument and their ($h_{5\mu},
h_{55}$)
compatibility conditions with the tensor $h_{\mn}$-equation.
It turns out  that
$h_{55}$ does not have any  bulk (massive) and brane (massless) propagations.
Although $h_{5\mu}$ has a sort  of  massive
propagations, they do not belong to the physical solution.
Hence we confirm that the Randall-Sundrum gauge suffices
the on-shell brane physics.

\end{abstract}

\newpage
\section{Introduction}

Recently there has been much interest in the Randall-Sundrum
brane world\cite{RS1,RS2}.
The key idea of this model is that our universe may be a brane embedded in
higher dimensional space. A concrete model is a single 3-brane embedded
in five-dimensional anti-de Sitter space (${\rm AdS}_5$). Randall and
Sundrum have shown that the longitudinal  part ($h_{\mu\nu}$) of the
metric fluctuation satisfies the Schr\"odinger-like equation with an
attractive delta-function\cite{RS2}. As a result, the massless zero mode which
describes the localized gravity on the brane was found.

However, we wish to point out that this has been done with the RS gauge
(a 4D transverse-tracefree gauge with $h_{5\mu} =h_{55}=0$).
It seems that this choice may be so restrictive because under this gauge
 the RS model can
describe the tensor fluctuation only.
In the Kaluza-Klein approach $h_{55}$ is a 4D scalar and $h_{5\mu}$ is
a 4D vector. Hence
 it is  natural to set these
fields to be non-zero at the beginning.
In order to
study the well-defined theory on the brane, it would be better to include  the
non-zero transverse parts of $h_{5\mu}, h_{55}$.  Ivanov and
Volovich\cite{IV}
found that the equation for $h_{55}$ includes a repulsive delta-function.
Also the equation of $h_{5\mu}$ has an attractive delta-function potential\cite{MKL}
\footnote{But this can be gauged away. See ref.\cite{MK}.}.

In this letter, we analyze  all the linearized equations including
$h_{5\mu}, h_{55}$ without introducing any matter source.
Hence we study the on-shell propagation of their metric fluctuations.
We choose initially the  de Donder gauge instead of the RS gauge.
 Consequently we will prove the no-go theorem which states that  $h_{5\mu}$ and $h_{55}$
do not have any  propagation  on the brane.

We start from the second RS model with a positive 3-brane at $z=0$\footnote{If the localized
matter of $T_{\mn}(x,z)=\delta(z)T_{\mn}(x)$ is introduced on the brane,
the brane is shifted from $z=0$ to $z=\ell$\cite{CNW}.} in
AdS$_5$\cite{RS2,IV,MKL}
\beq
\hat R_{MN}-\fr{1}{2} \hat g_{MN} \hat R = \Lambda
\hat g_{MN} +\sigma \fr{\sqrt{-\hat g}}{\sqrt{-\hat g_B}} \hat g_{B \mu\nu}
\delta^{\mu}_M \delta^{\nu}_N,
\label{EEQ}
\eeq
which is derived from the action
\beq
I = \fr{1}{2} \int d^5x \sqrt{-\hat g} (\hat R+2\Lambda )
 +\tilde\sigma\int d^4x \sqrt{-\hat g_B}
\label{Action}
\eeq
with $\kappa^2_5= 8 \pi G_5 =1$.
The RS solution is given in terms of  the conformal coordinates $(x,z)$ as

\beq
ds^2_{RS}=\hat g_{MN}dx^Mdx^N = H^{-2} g_{MN}dx^Mdx^N
\label{RSB}
\eeq
with $H=k|z|+1$ and $ g_{MN}=\eta_{MN} = {\rm diag}(+----)$.
 For $\eta_{MN} = {\rm diag}(-++++)$ convention, see ref.\cite{KM1}.
 In this solution
the bulk cosmological constant $\Lambda = -6k^2$ is fine-tuned by
the  brane tension $\tilde\sigma = 6k$.
Here the capital
indices $M,N, \cdots$ are split into $\mu, \nu, \cdots $
(four-dimensions: $x^{\mu}=x$) and $5(x^5=z)$.

For  convenience, it is important to use the conformal transformation of
$\hat g_{MN} = \Omega^2 { g}_{MN}$ with $\Omega = H^{-1}$.
Introducing a perturbation for the canonical metric
$g_{MN} = \eta_{MN} +h_{MN}$, then the line element takes the form
\beq
ds^2_{RSp}=H^{-2}\Big \{ (\eta_{\mn} + h_{\mn})dx^\mu dx^\nu
+ 2h_{5\mu}d zdx^\mu +(-1 +h_{55})dz^2 \Big \}.
\label{Fluctuation}
\eeq
Before we proceed, it is necessary to comment on the Z$_2$-symmetry
argument. We will use this as a guideline for our analysis.
As is easily checked in Eq.~(\ref{RSB}),
this is based on the fact that RS background is symmetric under $z \to -z$.
We require that this orbifold symmetry be preserved up to the linearized level
for the consistency.
In order that $ds^2_{RSp}$
have this symmetry, it requires that $h_{\mn}(x,z),h_{55}(x,z)$
be even with respect to $z$, but $h_{5\mu}$   be odd
: $h_{5\mu}(x,-z)=-h_{5\mu}(x,z)$.
This implies that $h_{5\mu}(x,0)=0$ on the brane. In other words, we can
gauge it away for the brane physics.
And  $h_{55}$ does not vanish in general. However, this is
not the whole story. Actually the propagations around the given
background must  be determined by solving their linearized equations thoroughly.
These equations can be obtained either from a linearized process of the
equation (\ref{EEQ}) or a  variational  process to the bilinear action which can be obtained
from Eq.(\ref{Action}).
Here we choose the former one because the  latter seems to be a
difficult process.

The linearized equation for
Eq.(\ref{EEQ}) takes the form\cite{MKL}
\beqa
\mbox{} && \Box h_{MN} +3\fr{\partial_K H}{H} \eta^{KL}
\left (\partial_N h_{KM} +\partial_M h_{KN} -\partial_K h_{MN} \right )
\nonumber  \\
\mbox{} && -\left (\fr{2\Lambda}{H^2} + \fr{2\sigma}{H}\right ) h_{55}\eta_{MN}
-\fr{2\sigma}{H}  \Big\{h_{MN}
- \left(h_{\mu\nu} + \fr{h_{55}}{2} \eta_{\mn} \right)\delta^{\mu}_M \delta^{\nu}_N \Big\} =0.
\label{FEQ}
\eeqa
We stress again that the above equation is suitable for describing
the vacuum propagation of metric fluctuations because we do not
introduce any localized matter source on the brane and matter source in
the bulk. For a localized matter and the RS gauge, see ref.\cite{GT,GKR}. For a bulk
matter source, their propagation was discussed in ref.\cite{KM1}.
For our purpose,  we choose the de Donder gauge
\beq
\partial^M h_{MN} =0, \qquad \qquad h^P_P=0,
\label{Donder1}
\eeq
which means that
\beq
h^{\mu}_{\mu} = h_{55}, \qquad \partial^{\mu} h_{\mu 5} =
\partial_5 h_{55}, \qquad \partial^{\mu} h_{\mu\nu} =
\partial_5 h_{5\nu}.
\label{Donder2}
\eeq
The 5D harmonic gauge with $h^P_P \not=0$ may be useful for the
study of the brane world with the bulk matter source\cite{KM1}.
 But this gauge is not suitable here. This is  because under this gauge, it is not easy to
diagonalize the linearized equations to obtain the eigenmodes.
 From Eq.(\ref{FEQ}) we obtain three equations
\begin{eqnarray}
&& \left ( \Box - {{12 k^2} \over H^2} - 3 f \partial_z \right )
             h_{55} =0 ,
\label{eqh55} \\
&& \left ( \Box - {{12 k} \over H} \delta(z) \right ) h_{5\mu}
   - 3 f \partial_\mu h_{55} =0 ,
\label{eqh5mu} \\
&& \left ( \Box + 3 f \partial_z \right ) h_{\mu\nu}
  - 3 f \left ( \partial_\mu h_{5\nu} +\partial_\nu h_{5\mu} \right )
  + {12k \over H} \left ( \fr{k}{H} - \fr{\delta(z)}{2}\right ) h_{55}\eta_{\mu\nu} =0
\label{eqhmunu}
\end{eqnarray}
with $f = \partial_z H/H=H'/H$.

\section{Graviscalar propagation}
First  let us study the  propagation of  $h_{55}$.
In analyzing the perturbation, if one encounters  a decoupled equation like Eq.(\ref{eqh55}),
one should solve it first. One  usually transforms it into
the Schr\"odinger-like equation to get an intuitive understanding.
And then one has to check its consistency with the
remaining equations (\ref{eqh5mu}) and (\ref{eqhmunu}).
Let us assume $h_{55}(x,z)= H^p(z) \psi_5(z)
\hat h_{55}(x)$\cite{IV}.  For $p=-3/2$, Eq.~(\ref{eqh55}) takes the
form
\begin{eqnarray}
&& \left ( \Box_4 + m_5^2 \right ) \hat h_{55}(x) =0,
\label{eqh55h} \\
&& - \fr{1}{2}\psi_5''(z) + V_5(z) \psi_5(z) =
              \fr{1}{2} m_5^2 \psi_5(z)
\label{eqpsi5}
\end{eqnarray}
with its potential
\beq
V_5(z)= \fr{3}{2} \fr{k}{H} \delta(z)
              - \fr{45}{8} \fr{k^2}{H^2} .
\label{eqpsi5V}
\eeq
We observe that an repulsive delta-function implies
the non-existence of   localized zero-mode on the brane. To check this thoroughly,
 we consider  $\psi^0_5=c_5 H^q(z)$ when $m_5^2=0$.
With $q=3/2$, Eq.~(\ref{eqpsi5}) leads to $\psi^0_5=0$.
Hence we do not find any zero mode for a type of $h_{55}^0= c_5 \hat
h_{55}(x)$ with constant $c_5$. This result may be related to
those for the  zero-mode approach\cite{ML,KM2,Myu}.
Another possibility is  that Eq.~(\ref{eqpsi5}) may give us a massive
 propagating solution along $z$-axis.
But considering its compatibility with tensor equation, the
answer is No. The tensor equation (\ref{eqhmunu}) plays
a  key role in our analysis because it contains all of
metric perturbations.
Taking the trace of Eq.~(\ref{eqhmunu}) and using the gauge condition of  Eq.~(\ref{Donder2})
leads to the other equation for $h_{55}$.
Comparing it with
Eq.~(\ref{eqh55}), one finds that $h_{55}=0$\cite{MKL}.
As a result, we prove the no-go theorem for the graviscalar\footnote{If there
is a delta function source on the brane, $h_{55}=0$ may break the five-dimensional
diffeomorphism and hence the coordinate $z$ becomes discontinuous across the brane.
In this case, it is reasonable to start with $h_{55}\not=0$\cite{Kak}.}.

\section{Gravivector propagation}

 Now we are in a position to discuss the gravivector propagation.
According to the no-go theorem for the gravivector,
one does not expect to find any vector  propagation on the brane\cite{RS1,LP}.
 This is based on the Z$_2$-symmetry
argument.  Now let us prove this
using the linearized equations.
Considering $h_{55}=0$, Eq.~(\ref{eqh5mu}) becomes a decoupled  vector equation.
Also  we have to solve this equation first too.
In order to solve Eq.~(\ref{eqh5mu}), we introduce the separation of
variables as
\begin{equation}
h_{5\mu}(x,z) =  \psi_v(z) \hat h_{5\mu}(x).
\label{separation}
\end{equation}
Then Eq.~(\ref{eqh5mu}) leads to
\begin{eqnarray}
&& \left ( \Box_4 + m_v^2 \right ) \hat h_{5\mu}(x)  =0,
\label{eqhmu} \\
&& - \fr{1}{2} \psi_v''(z) + V_v(z) \psi_v(z)
               = \fr{1}{2} m_v^2 \psi_v(z)
\label{eqpsiv}
\end{eqnarray}
with the attractive $\delta(z)$-type potential $V_v(z)$ as
\beq
V_v(z)= -{6 k \over H} \delta(z).
\label{eqhmuV}
\eeq
Now let us solve Eq.(\ref{eqpsiv}) first.
This is just the case of ref.\cite{Gri}. And then consider its
compatibility with the tensor equation.
Naively  we have two kinds of solutions : one bound state and scattering
states.

\subsection{Bound state solution with $m^2_v<0$}

This solution must satisfy the equation
$\psi_v''(z) + m_v^2 \psi_v(z)=0$ with $m_v^2=-\kappa^2 <0$ at everywhere, except $z=0$.
Its normalizable solution  is
\begin{equation}
\psi_v^{z<0}(z)= C e^{\kappa z}, ~~~~
\psi_v^{z>0}(z) = D e^{-\kappa z}.
\label{solpsiv1}
\end{equation}
It is expected that this solution has the $Z_2$-symmetry of the RS
solution  under  $z \to -z$\cite{RS2}.
This can be easily achieved here if $C=D$. Immediately one finds $C=D$
because of the continuity of the wave function at $z=0$.
The derivative of $\psi_v(z)$ is no longer continuous
due to the presence of the delta-function.
That is, one has
\begin{equation}
{{\partial \psi_v} \over {\partial z} } \Bigg |_{z=0^+}
-{{\partial \psi_v} \over {\partial z} } \Bigg |_{z=0^-}
= -12 k C,
\label{bc}
\end{equation}
which leads to
\begin{equation}
\kappa =  6k
\label{m5}
\end{equation}
This admits only one of the  normalizable bound state solution,
irrespective of whatever $k$ is, as
\beq
\psi_v^e(z)= \sqrt{6k} e^{-6k|z|}.
\label{solpsib}
\eeq
This is an even function with respect to $z$.
However, we do not accept this as a physical solution
because it gives us the tachyonic mass of $ m_v^2=- 36k^2$ in view of  Eq.~(\ref{eqhmu}).
Furthermore this solution does not satisfy the Z$_2$-symmetry
argument which dictates  that $h_{5\mu}$ is odd. Hence we exclude it.

\subsection{Scattering state solution with $m_v^2>0$}

In this case we start with a plane wave solution
\begin{equation}
\psi_v^{z<0}(z)= A e^{i m_v z}+ B e^{-im_v z}, ~~~~
\psi_v^{z>0}(z) = F e^{i m_v z} + G e^{-im_v z}.
\label{solpsiv2}
\end{equation}
The continuity of the wave function at $z=0$ requires
$A+B=F+G$, while the discontinuity of its derivative at $z=0$
gives $ im_v(F-G-A+B)=-12k(A+B)$. If this relation makes sense, one requests that
$A+B
\not=0$.
Note that we have five unknown quantities of $A,B,F,G,m_v$
but two equations in hand. In this case, at most,
we have $F= A+i\beta (A+B)$, $G=B- i\beta(A+B)$ with
$\beta=6k/m_v$. Especially, considering the incident wave propagation$(\to)$,
 we have $G=0$. This case leads to
a conventional scattering with the reflection coefficient ${\cal R}=\beta^2/(1+\beta^2)$
and the transmission coefficient ${\cal T}=1/(1+ \beta^2)$.

However it remains to check whether the above solution is or not
consistent with the tensor equation (\ref{eqhmunu}).
Acting $\partial^\mu$ on Eq.~(\ref{eqhmunu}) and using
Eqs.~(\ref{Donder2}) and (\ref{eqh5mu}),
one gets the compatibility condition with the tensor equation~\cite{MKL,Vol}
\begin{equation}
\left ( \delta(z) h_{5\mu} \right )'
- 3 {\rm sgn}(z) \delta(z) h_{5\mu} =0
\label{condition}
\end{equation}
with the step function of ${\rm sgn}(z)=\theta(z)$.
This is satisfied trivially at everywhere,  except $z=0$.
But  ${\rm sgn}(z)\delta(z)$ is not well defined at
$z=0$ and thus it requires
\begin{equation}
h_{5\mu}(x,0)=0 \to ~~\psi_v (0) =0.
\label{boundary}
\end{equation}
This is exactly the same requirement from the Z$_2$-symmetry argument.
Here  the plane
wave solution in Eq.~(\ref{solpsiv2}) does not satisfy the
above compatibility condition because of $A+B\not=0$. Therefore it cannot to be  a
solution.

\subsection{An alternative odd solution with $\psi_v(0)=0$}

Finally, we propose an alternative solution which may satisfy
Eqs.~(\ref{eqpsiv}) and (\ref{boundary}) simultaneously. This is
\begin{equation}
\psi^{z<0}_v (z) = 2iA \sin m_v z,~~~~
\psi^{z>0}_v(z) =2iF \sin m_v z.
\label{psivsol3}
\end{equation}
This can be obtained by demanding an additional condition $\psi_v(0)=0$
of
 Eq.~(\ref{solpsiv2}) : $ A+B=F+G=0 \to
A=-B,F=-G$. Further we have $A=F$ from  the discontinuity relation for the  derivative
of wave function. Then this leads to the propagation of  a free
particle,
\beq
\psi^o_v(z)=2iA \sin(m_v z).
\label{psivsca}
\eeq
We have a few comments on this solution in order.
First we get ${\cal T}=1,{\cal R}=0$, which means that it is not the scattering state.
Instead it satisfies  the boundary condition of $\psi^o_v(0)=0$
required by  both the compatibility and the Z$_2$-symmetry argument.
Also this is an odd function, as required by the symmetry argument.
However, importantly, we remind the reader that our background is AdS$_5$ with
the brane.
This means that the solution to the linearized equations can
carry at least the parameter `` $k$'' because the size of AdS$_5$-box
is $1/k$ approximately and the brane tension is $\tilde\sigma =6k$.
An example for this is the massive solution for
$h_{\mn}(x,z)$\cite{RS2}. Unfortunately  this  wave solution misses ``$k$''.
It seems that this corresponds to a free particle  propagating along $z$-axis.
 Due to the compatibility condition
(\ref{boundary}), this  solution
$\psi^o_v(z)$ does not account for the presence of the potential at
$z=0$ ($-6k\delta(z) \psi_v(z)$-term in Eq.~(\ref{eqpsiv})) appropriately.
On the other hand, the  bound state solution $\psi_v^e(z)$ in Eq.~(\ref{solpsib})
is obtained by taking into account
 $\delta(z)\psi_v(z)$ very well. However this does not satisfy
 the compatibility condition of $\psi_v(0)=0$. Also it generates the
 tachyonic mass for the gravivector.
 A solution
of Eq.~(\ref{psivsca}) is neither a scattering state nor a bound state.
Therefore it cannot be regarded as a truly massive propagating
solution which accounts for the RS background well.
It is obvious that there is no consistent  massive solution which satisfies both
Eqs.~(\ref{eqh5mu}) and (\ref{eqhmunu}).
Lastly, the massless
vector propagation is not allowed inherently  because with $m_v=0$, all
solutions (\ref{solpsib}), (\ref{solpsiv2}), and (\ref{psivsol3})
 imply that there is no wave along $z$-axis.
 Even if $m_v\not=0$, we cannot find any form of
  $h^o_{5\mu}(x,0)(\equiv \psi^o_v(0)\hat h_{5\mu}(x))= c_v \hat h_{5\mu}(x)$ because
 of $\psi^o_v(0)=0$.

 All of these results
 support the symmetry argument for the non-existence of
  massless gravivector on the brane\cite{RS1,LP}.  $h_{5\mu}$
  is considered as  the gauge degrees of freedom on the
  brane\cite{MK}.
  Hence if one encounters $h_{5\mu}$, it can  be always gauged  away
  without loss of generality.

\section{Graviton propagation}

 Finally  the tensor equation (\ref{eqhmunu}) with $h_{55} =h_{5\mu} =0$ becomes
\beq
(\Box +3f\partial_5) h_{\mu\nu} =0,
\eeq
which just corresponds to the RS  case.
Introducing the  variables
$h_{\mu\nu} = H^{3/2}(z)
\psi_h(z) \hat h_{\mu\nu}(x)$, one finds
\beq
(\Box_4 +m^2_h)\hat{h}_{\mu\nu} =0,~~~
 -\fr{1}{2} \psi_h'' + V_h(z)  \psi_h (z)
= \fr{1}{2} m^2_h \psi_h(z)
\eeq
with the potential
\beq
V_h(z) = \fr{15}{8} \fr{k^2}{H^2} -\fr{3}{2} \fr{k}{H} \delta (z).
\eeq
Using the volcano-type potential $V_h(r)$, it is easily to prove  $m^2_h \ge 0$.
This fact implies that these are no normalizable bound states for
graviton modes. Also the attractive $\delta(z)$-term guarantees the localized
zero mode solution $h^0_{\mn}=c_h \hat h_{\mn}(x)$ for gravitons at $z=0$\cite{RS2}.
\section{Discussions}

 We have proved  that
 the gravivector $(h_{5\mu})$ and graviscalar $(h_{55})$
cannot have any massless mode on the vacuum RS brane.
In addition the graviscalar does not have any massive propagation
along $z$-axis. In the case of gravivector, we have a sort of
massive propagations.
But these cannot be a candidate of  the physical solution because
they cannot account for the background situation
correctly.  Our result is based on the Z$_2$-symmetry argument and their
compatibility conditions with the tensor $h_{\mn}$-equation.
The first is  a guideline for our analysis,
whereas the second plays a role as a criterion for
selecting a physical solution. Two require the
same condition for the gravivector.
On the other hand,  for the graviscalar, the first implies
only that it must be an even function but the second gives us the other equation for $h_{55}$.
 Comparing
it with Eq.~(\ref{eqh55}) leads to $h_{55}=0$. Thus the compatibility condition for
$h_{55}$ provides a very strong constraint for the non-propagation of
the graviscalar on the brane.

Up to now we do not consider the matter source. We comment on the inclusion of matter.
 There exists a brane-bending effect with the
RS gauge which leads to the other massless scalar by introducing the localized
matter source on the brane\cite{GT,GKR}.
Here the coordinates are discontinuous because the brane appears
"bent" in the presence of a localized matter on the brane.
Recently, authors in \cite{Kak,AIMVV} carried out the same problem
but with alternative gauges for which the coordinates are
continuous across the brane.

In conclusion, it is explicitly proved that there is no on-shell propagations
for the graviscalar and gravivector. This means that the RS gauge
($\partial^\mu h_{\mn}=h=h_{5\mu}=h_{55}=0$)
suffices the on-shell brane physics when  any matter is absent\footnote{
But Tanaka\cite{TAN} insisted that the RS gauge is general even if the matter is absent.
 Hence, starting with the de Donder gauge, there are still remaining gauge
degrees of freedom which eliminate all scalar and vector modes. This does not
contradict with our result. In our analysis, if one consider the on-shell degrees of
freedom as the physical ones, our mode of Eq.(\ref{psivsca}) corresponds to a gauge
degree of freedom (an off-shell mode).}.
Even if one introduces a localized matter on the brane, it is
reasonable to choose the axial gauge $h_{5\mu}=0$. This is so
because it belongs to the gauge degrees of freedom. But the gauge
choice of $h_{55}=0$ is not still justified when the localized matter is
present\cite{Kak,CNW}.

\section*{Acknowledgments}

We would like to thank Gungwon Kang,  H. W. Lee, especially  Ishwaree Neupane  for helpful
discussions.
This work was supported by the Brain Korea 21 Program, Ministry of
Education, Project No. D-0025 and KOSEF, Project No.2000-1-11200-001-3.

\end{document}